\documentclass[12pt,epsfig]{article}
\usepackage{graphicx,subfigure,amsmath,latexsym,amssymb, mathtools}
\usepackage{epsfig,pdfpages,float,changepage}
\usepackage{natbib, longtable}
\usepackage{rotating}
\usepackage{braket}
\usepackage{tikz}
\usepackage{authblk}
\usepackage{esvect}

\DeclareMathOperator*{\argmax}{arg\,max}

\providecommand{\keywords}[1]
{
  \small	
  \textbf{\textit{Keywords---}} #1
}

\begin{document}


\title{Applications of Quantum Annealing in Statistics}
\author{Robert C. Foster, Brian Weaver, Jim Gattiker}
\affil{Los Alamos National Laboratory}

\maketitle

\begin{abstract}

Quantum computation offers exciting new possibilities for statistics. This paper explores the use of the D-Wave machine, a specialized type of quantum computer, which performs quantum annealing. A general description of quantum annealing through the use of the D-Wave is given, along with technical issues to be encountered. Quantum annealing is used to perform maximum likelihood estimation, generate an experimental design, and perform matrix inversion. Though the results show that quantum computing is still at an early stage which is not yet superior to classical computation, there is promise for quantum computation in the future.

\end{abstract}

\\

\vspace{0.75in}

\begin{centering}

\large LA-UR-19-23207

\end{centering}


\cleardoublepage



\section{Introduction}


Quantum computing has arrived. Long imagined as a technology of the future, early quantum computers exist today which are being used to solve real problems. These computers, however, exist in what has been dubbed the noisy intermediate-scale quantum (NISQ) era (\cite{Preskill2018}), with enough qubits that classical computers can not simply simulate the machines, but with hardware issues introducing noise into the computation and an inability to perform error-checking.


This paper will focus on one particular type of quantum computer - the D-Wave quantum annealer. It must be stressed that this is not a universal quantum computer which can be programmed to run various types of quantum computing algorithms, but rather a specialty quantum computer built to solve combinatorial optimization problems using one type of quantum computing algorithm. The advantage of focusing on the D-Wave is two-fold: first, that quantum annealing using the D-Wave is much simpler to perform and understand than algorithms which run on universal quantum computers, which often requires knowledge of quantum mechanics to comprehend and whose programming requires direct implementation of quantum logic gates. Second, it is currently much more readily available for use through companies and organizations such as the Universities Space Research Association (USRA). Quantum annealing may potentially provide a polynomial increase in computing power depending on the problem at hand, though research is ongoing. 

The purpose of this paper is not to give an in-depth discussion into the underlying physics of quantum computation or hardware of the D-Wave quantum annealer, but rather to provide an introduction to quantum annealing and the D-Wave hardware in general, and suggest directions for future research. The original paper discussing the use of the Ising model for quantum annealing is \cite{Bian2010}. An excellent overview of quantum annealing may be found in \cite{Biswas2017}.  Within the statistical literature, \cite{Wang2016} also provides a thorough description of both the physics and hardware of the D-Wave machine, and derives statistical theory for tests comparing the results of D-Wave output to results from classical algorithms. Section \ref{Sec:QuantumBits} gives a general description of quantum bits, focusing on the property needed to understand quantum annealing, and a general description of quantum annealing through the adiabatic theorem. Section \ref{Sec:Technical} describes technical issues which affect the performance of the quantum annealing algorithm on the D-Wave. Sections \ref{Sec:MaximumLikelihood}, \ref{Sec:NQueens}, \ref{Sec:MatrixInversion} give examples of quantum annealing using the D-Wave for maximum likelihood, experimental design generation, and matrix inversion, respectively.

\section{Quantum Computation using the D-Wave}
\label{Sec:QuantumBits}

Before beginning a discussion of what a quantum computer is, it is important to clarify what it is not. A quantum computer is not simply a traditional computer that operates with much more computational power. In fact, for most problems traditional computation is faster. If all that existed were quantum computers, it would still be necessary to invent traditional computers in order to improve performance for most applications. And a quantum computer is not simply a machine that operates on all answers simultaneously and produces the correct answer as if by magic - its mathematics are much more subtle.

\subsection{Quantum Bits}

A quantum computer is a device which exploits properties of elementary particles to perform linear algebra in the complex plane without the requirement of storing numerical values in memory and performing computational operations on them. If a traditional bit may only take the values 0 and 1, define a qubit, short for quantum bit, to be a particle existing in the form

\begin{equation}
q = c_0 \ket{0} + c_1 \ket{1}
\label{Eq:Qubit}
\end{equation}

\noindent where $c_0, c_1 \in \mathbb{C}$ and $\ket{0}, \ket{1}$ are, using standard bra-ket notation, orthonormal basis vectors in the complex plane corresponding to physical properties of the particle such as spin but nominally representing numerical values. The complex coefficients $c_0$ and $c_1$ must have $|c_0| + |c_1| = 1$, and with this constraint the set of values that $q$ may take is often represented as a sphere in complex space. In this way, the qubit manages to exist as a  linear combination of the vectors $\ket{0}$ and $\ket{1}$. The qubit is connected to the traditional bit by associating the pure $\ket{0}$ and pure $\ket{1}$ states with numerical values. These values may theoretically be anything, but most common are $\{-1, 1\}$ and $\{0, 1\}$. For this paper, define 0 be the numerical value associated with the superposition defined by $c_0 = 1$ and $c_1 = 0$, and likewise 1 as the numerical value associated with the superposition defined by $c_0 = 0$ and $c_1 = 1$. Given values of $c_0$ and $c_1$ before observation of the qubit, upon observation the qubit $q$ collapses into either 0 or 1 with probability $P(0) = |c_0|$ and $P(1) = |c_1|$. Further properties of quantum mechanics are necessary for other quantum computing algorithms but the superposition property is all that is needed for quantum annealing.








\subsection{Problem Format}

The D-Wave is not a universal quantum computer. It is strictly an optimization machine which solves one specific problem. The D-Wave in principle finds the set of $q_i$ that minimize the energy function

\begin{equation}
Energy = \sum_i a_i q_i + \sum_{j > i} b_{ij} q_i q_j
\label{Eq:Energy}
\end{equation}

\noindent where the $q_i$ are binary variables that can fit into one of two forms: either $q_i \in \{-1, 1\}$, known as the Ising model, or $q_i \in \{0, 1\}$, known as the quadratic unconstrained binary optimization (QUBO) model. The $a_i$ and $b_{ij}$ can be any real number, though the physical properties of the D-Wave hardware impose minimum and maximum values  for each and rescaling must be applied in the case that $a_i$ and $b_{ij}$ values input by the user are outside this range. For a system with $n_q$ qubits, there are $2^{n_q}$ possible solutions. This exponential scaling of possible solutions means the computation required to find the minimum energy of systems as in Equation (\ref{Eq:Energy}) also scales exponentially as the problem size increases.

The D-Wave operates natively using the Ising model, and any QUBO models input to the machine are converted to Ising models before running. This paper will work primarily with the QUBO model, as it is much more natural for the problems considered. A QUBO problem can also be written in the form

\begin{equation}
Energy = x' Q x
\label{Eq:QUBOMatrix}
\end{equation}

\noindent where $x$ is an $n_q$ length vector consisting of the $q_i$ values and $Q$ is an upper-triangular matrix consisting of the $a_i$ values along the diagonals and the $b_{ij}$ values in the upper triangle. Each of the two models, Ising and QUBO, may be converted back and forth to each other using linear transformations, potentially including linear offsets, and D-Wave provides automated software to perform this. Formatting a problem to run in QUBO form on the D-Wave means determining a matrix $Q$ as in Equation (\ref{Eq:QUBOMatrix}). 

\begin{figure} \centering
\includegraphics[height = 50mm]{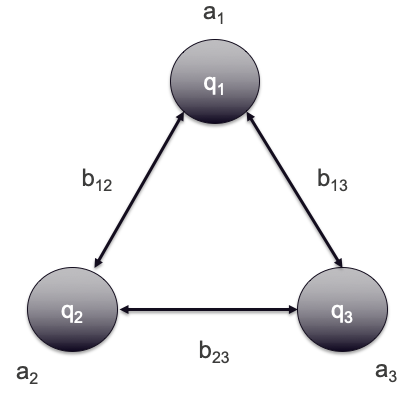} 
\includegraphics[height = 40mm]{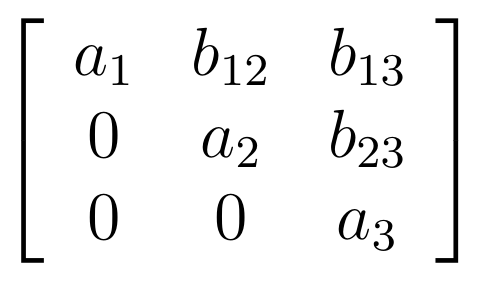} 
\caption{A simple three-qubit system. Each qubit has weight $a_i$ and each pair of qubits is connected with interaction $b_{ij}$. Each qubit may take one of two values: either $q_i \in \{-1, 1\}$, known as the Ising model, or  $q_i \in \{0, 1\}$, known as the QUBO model. The energy of the system is given by plugging the $q_i$ values into Equation (\ref{Eq:Energy}). When the $q_i$ follow the QUBO model, the system can be represented by the upper-triangular matrix on the right, and the energy calculated by Equation (\ref{Eq:QUBOMatrix}).}
\label{Fig:ThreeQubitSystem}
\end{figure}

Each system as in Equation (\ref{Eq:Energy}) can also be thought of a graphical network between the $n_q$ qubits, wherein each qubit is assigned weight $a_i$ to itself and any interactions between qubits are assigned weight $b_{ij}$. This visualization as a graphical network is important to the operation of the D-Wave machine as the graph must be mapped onto the physical hardware, which will be discussed in Section \ref{Sec:Technical}. A three-qubit system and corresponding QUBO matrix are shown in Figure \ref{Fig:ThreeQubitSystem}.

\subsection{Quantum Annealing}

Systems in the form of Equation (\ref{Eq:Energy}) are commonly optimized by techniques such as simulated annealing which are able to both climb and descend peaks in the energy surface in order to avoid becoming trapped in local optima. In the worst case, however, they still may require an enumeration of all $2^{n_q}$ possible combinations of qubits in order to find the lowest-energy solution. The quantum annealing procedure, in theory, allows for direct estimation of the minimum energy state without traversal of the entire energy surface. Maintaining quantum coherence of the qubit in a linear combination of 0 and 1 states, as in Equation (\ref{Eq:Qubit}), allows for a phenomenon called quantum tunneling, in which the solution travels directly through an energy barrier between local minima without the ``backing-out'' of a local optima that a procedure like simulated annealing performs. 



\begin{figure} \centering
\includegraphics[height = 100mm]{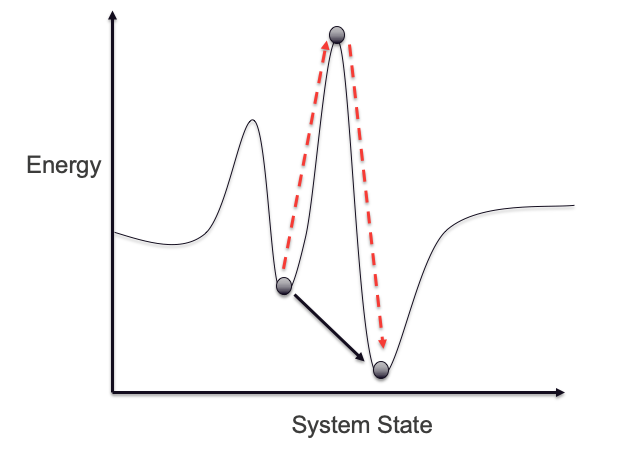} 
\caption{During simulated annealing, indicated by the dashed line, the algorithm must climb hills in the energy surface to escape local minima and traverse to other potential solutions. Quantum annealing, indicated by the solid line, allows for direct traversal between minima by tunneling through hills in the energy surface. This relies on maintaining quantum coherence of the qubits, as in Equation (\ref{Eq:Qubit}). Quantum annealing should theoretically outperform simulated annealing on energy surfaces which feature tall, thin peaks in the energy surface. Non-artificial problems in which quantum annealing outperforms simulated annealing are difficult to find.}
\label{Fig:EnergySurface}
\end{figure}

Figure \ref{Fig:EnergySurface} demonstrates the manner in which a solution moves through the optima in the energy surface in order to find the global optima. In theory, quantum annealing performs best on energy surfaces with tall, narrow peaks that impose substantial difficulties on hill-climbing techniques, but require only a short tunnel for the quantum anneal. In practice, such problems are hard to find. Attempts to construct problems which show supremacy of the quantum annealer over classical techniques, have so far been focused on artificially constructed problems,  as described in \cite{Mandra2017}.

The D-Wave performs this quantum annealing through the use of the adiabatic theorem. The adiabatic theorem, stated simply, says that if a system begins in the minimum energy state of a Hamiltonian, where a Hamiltonian is a system of qubits, weights, and interactions as in Equation (\ref{Eq:Energy}) and Figure \ref{Fig:ThreeQubitSystem}, and transitions to a new Hamiltonian slowly enough, then the system remains in the lowest-energy state the entire time. This is, of course, a vast simplification of the procedure. A more complete technical description, including the underlying physics, may be found in \cite{Biswas2017}.

\begin{figure} \centering
\includegraphics[height=100mm]{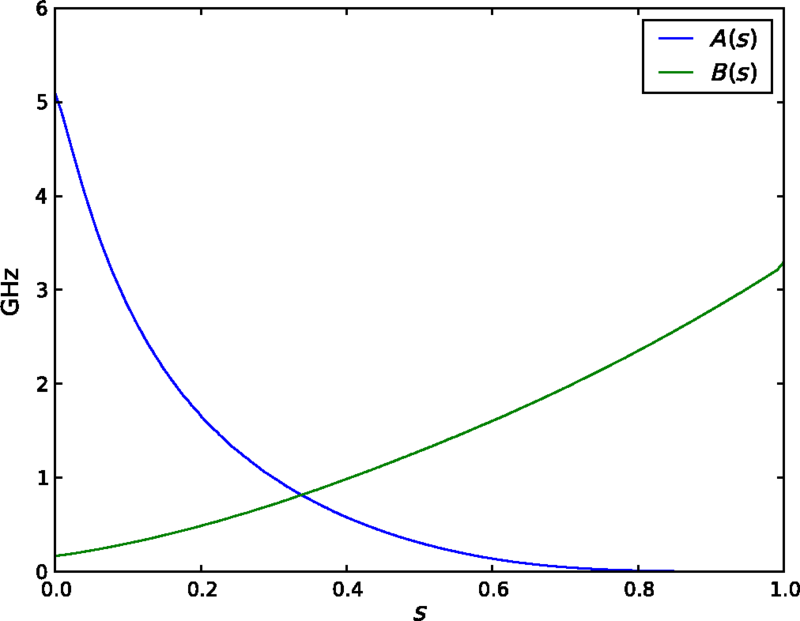} 
\caption{The basic annealing schedule for the D-Wave quantum annealer, where $A(s)$ and $B(s)$ are as in Equation (\ref{Eq:Anneal}). At $s = 0$, $A(s) > 0$ and $B(s) = 0$, so the D-Wave is set entirely to system $H_0$ which has an easily located lowest-energy state. At $s = 1$, $A(s) = 0$ and $B(s) > 0$, so the D-Wave is set entirely to the user-input system $H_1$ as in Equation (\ref{Eq:Energy}). If the annealing is done slowly enough, the adiabatic theorem states that the system will stay in the lowest-energy configuration of $q_i$ the entire time. The user has control over the annealing time and other aspects of the anneal. Figure originally from \cite{Katzgraber2015}.}
\label{Fig:AnnealingSchedule}
\end{figure}

Beginning with a predetermined Hamiltonian $H_0$ in which the lowest-energy state is easily found, the system transitions to the user-input Hamiltonian $H_1$ through the Equation

\begin{equation}
H(s) = A(s) H_0 + B(s) H_1
\label{Eq:Anneal}
\end{equation}

\noindent where $s$ transitions smoothly from $0$ to $1$, and $A(s)$ are chosen such that $A(0) > 0$ and $B(1) = 0$ at the beginning of the anneal, while $A(1) = 0$ and $B(1) > 0$ at the end of the anneal. This is shown in Figure \ref{Fig:AnnealingSchedule}. By changing the time it takes to transition $s$ from $0$ to $1$, called the annealing schedule, the user can determine the amount time it takes to complete the anneal in Equation (\ref{Eq:Anneal}). The minimum annealing time is 10 $\mu s$. Whereas in simulated annealing the energy system remains constant and bits are changed in order to explore it, in quantum annealing the energy system is changed around the qubits and physics forces them to adapt in order to remain in the lowest-energy state.

How slowly is slowly enough? The system is only guaranteed to arrive at the lowest-energy state in the case of an anneal of infinite length time. In general, the necessary anneal time increases with the complexity of the energy surface. The required annealing time is determined by the energy gap between the lowest-energy solution and the second lowest-energy solution at any point during the anneal process, at which the qubits are likely to tunnel from the optimal solution to a non-optimal solution. It is currently unknown whether or not this gap decreases exponentially as problem size grows, though if the system could be run at zero temperature it is known that this gap would decrease polynomially. A polynomial decrease in computational time, however, would still yield a sizeable increase in computational power, taking an algorithm which runs in cubic time to square time or lower, for example, or an algorithm which runs in linear time to square root time as with Grover's search algorithm (\cite{Grover1996}). Some evidence suggests that the temperature may need to drop at minimum in a logarithmic rate, or possibly at a power rate, as the problem size increases (\cite{Albash2017}).

Because the system is not guaranteed to stay in the lowest-energy state for a given anneal, it is common to perform a large number of anneals and return the set of $q_i$ and energy for each. The default number of anneals is currently 1000 but can be increased to up to 10000 by the user. The machine will usually return some low-energy solution for each anneal, and so over a large number of samples there is a probability of returning the lowest-energy solution which should be high for well-formed problems. The user then takes the solution with the lowest-energy state as the final result. It is entirely possible, however, that the machine will fail to find the lowest-energy solution. Because of this, users should not attempt to use the D-Wave to solve problems in which obtaining only a good solution rather than the best solution is disastrous. 

\cite{Bian2010} notes that under theoretically optimal operating conditions, the distribution of outcomes can be modeled as a Boltzmann distribution.

\begin{equation}
P(\{q_1, q_2, \dots\}) \propto \exp\left(\dfrac{ \sum_i a_i q_i + \sum_{j > i} b_{ij} q_i q_j}{\tau}\right)
\label{Eq:Boltzmann}
\end{equation}

\noindent In Equation (\ref{Eq:Boltzmann}), the temperature parameter $\tau$ can be thought of as corresponding to physical thermal effects present in the hardware due to the machine running at nonzero temperature. A number of estimators for the value of $\tau$ have been developed, as discussed in \cite{Raymond2016}. Operating conditions are not optimal, however, and empirical output from the machine shows significant deviations from the model in Equation (\ref{Eq:Boltzmann}). 

\subsection{General Hardware Criticisms}

The D-Wave quantum computer is, to put it mildly, a controversial machine within the quantum computing community, with some questioning how ``quantum'' the machine really is (\cite{Shin2014}). All types of quantum computing rely upon maintaining the coherence of quantum particles as in Equation (\ref{Eq:Qubit}), and the lengthening time until decoherence continues to prove a formidable problem. The decoherence time of qubits in the D-Wave is reported to be on the order of nanoseconds. The annealing time is at a minimum of 1 $\mu s$, which is orders of magnitude longer than the decoherence time. The initial question regarding the D-Wave hardware, then, was whether the machine was performing any sort of quantum computation at all. This was answered affirmatively in \cite{Johnson2011}, which compared the results of quantum annealing using the D-Wave to classical techniques and showed that the D-Wave output exhibited characteristics that could only have occurred with quantum tunneling. The next question was whether any problem could be formulated, even artificially, which showed quantum supremacy over the best classical algorithms. Though quantum supremacy as a concept is still ill-defined (\cite{Ronnow2014}), attempts are currently being made to obtain it using the D-Wave (though not without controversy) as in \cite{Mandra2017}. The final question remains whether any quantum supremacy the D-Wave shows on the Ising spin energy minimizations is because of its quantum computational nature, or because it is a machine specifically built for solving  Ising spin energy minimizations. All hope is not lost, however - future improvements of the machine may allow for demonstration of quantum supremacy strictly due to quantum effects, and since the D-Wave is known to perform quantum tunneling, however briefly, it is still worthwhile to explore the computational model.


\section{Hardware and Technical Issues Affecting the Anneal}
\label{Sec:Technical}

All computers, quantum or otherwise, must map a problem into a format the computer can understand. In the classical realm, this is using binary floating point arithmetic to represent numbers as bits, for example. These issues appeared complex in early classical computing, but the effect of these classical technical issues has been reduced to a level that it can be safely ignored for the vast majority of calculations. Quantum computers exhibit technical issues analogous to those faced by early classical computers, however, and as such are currently both limited in the problems they can represent and noisy in the ways they solve these problems.

\subsection{Rescaling and Added Noise}
\label{Sec:AddedNoise}

As previously mentioned, the range of values allowed for the $a_i$ and $b_{ij}$ in Equation (\ref{Eq:Energy}) is limited in the hardware implementation of the algorithm. On the D-Wave 2000Q, these values are $a_i \in [-2.0, 2.0]$ and $b_{ij} \in [-4.0, 1.0]$. These values are for the Ising model, not the QUBO model, and so care must be taken when implementing algorithms to ensure that $a_i$ and $b_{ij}$ values which are reasonable in the QUBO model do not become wildly out of range of the hardware in the resulting Ising model. If there are $a_i$ or $b_{ij}$ values outside of this range, the entire set of coefficients must be rescaled so that all the $a_i$ and $b_{ij}$ fit within the range of the hardware.


A rescaling of the coefficients would not be a problem, being equivalent to multiplying both sides of Equation (\ref{Eq:Energy}) by a constant, except that the values of $a_i$ and $b_{ij}$ are themselves subject to noise. This noise is due to both imperfections in the hardware and thermal fluctuations which affect the physical implementation of the annealing process. A model for the noisy coefficients is

\begin{equation}
\begin{aligned}
a_i^* = a _i + \epsilon\\
b_{ij}^* = b_{ij} + \delta
\end{aligned}
\label{Eq:Noise}
\end{equation}

\noindent where $\epsilon$ and $\delta$ are random variables, ideally independent between qubits, but evidence of both temporal and spatial correlation of errors exists (\cite{Michalak2017}). These have been commonly modeled as mean-zero Gaussian random variables, though \cite{PerdomoOrtiz2016} gives empirical evidence of bias, and also offers a method of bias correction. Intuitively, the values of $a_i$ and $b_{ij}$ which are smallest in magnitude are those most likely to be affected by the noise. In general, these errors can not be avoided. Decreasing the standard deviations of the noise components will likely involve lowering the operating temperature of the D-Wave machine (\cite{Albash2017}) - not an easy task, as it already runs at $0.015$ degrees Kelvin. The power draw necessary for this cooling is claimed to be less than 25 kW,  far less than the US Department of Energy's stated goal of 20-30 MW power for an exascale supercomputer (\cite{King2017}).

These errors can both significantly alter the system so that the lowest-energy state is no longer the one desired by the user and degrade the performance of quantum annealing so that it shows no speedup over classical algorithms (\cite{Venturelli2015}). These errors also prevent the D-Wave from being used for true random number generation, as even if the zero model is used with all $a_i = b_{ij} = 0$ so that every set of $q_i$ is theoretically equally likely,  the results of the qubit spins are still not independent Bernoulli random variables.  Research is ongoing into models for the D-Wave output which incorporates noise. \cite{Coffrin2019} gives an improved statistical model over the Boltzmann sampler of Equation (\ref{Eq:Boltzmann}) for the output from the D-Wave as a hierarchical model wherein the $\epsilon$ in Equation (\ref{Eq:Noise}) are modeled as mean zero Gaussians, and gives evidence that this model provides a closer match of output from the D-Wave hardware to theoretical probabilities. This is only an approximate model, however, and there are still sources of noise beyond errors in coefficients which causes the output from the D-Wave to deviate from the noisy Boltzmann sampler model.



\noindent 

\subsection{Embedding the Problem Graph}

\begin{figure}  \centering
\includegraphics[height = 50mm]{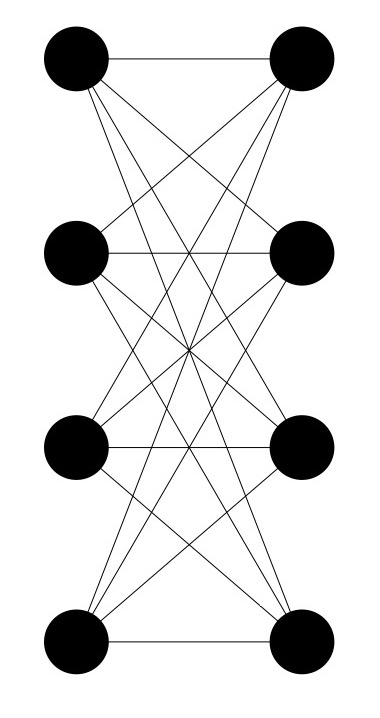} 
\includegraphics[height = 60mm]{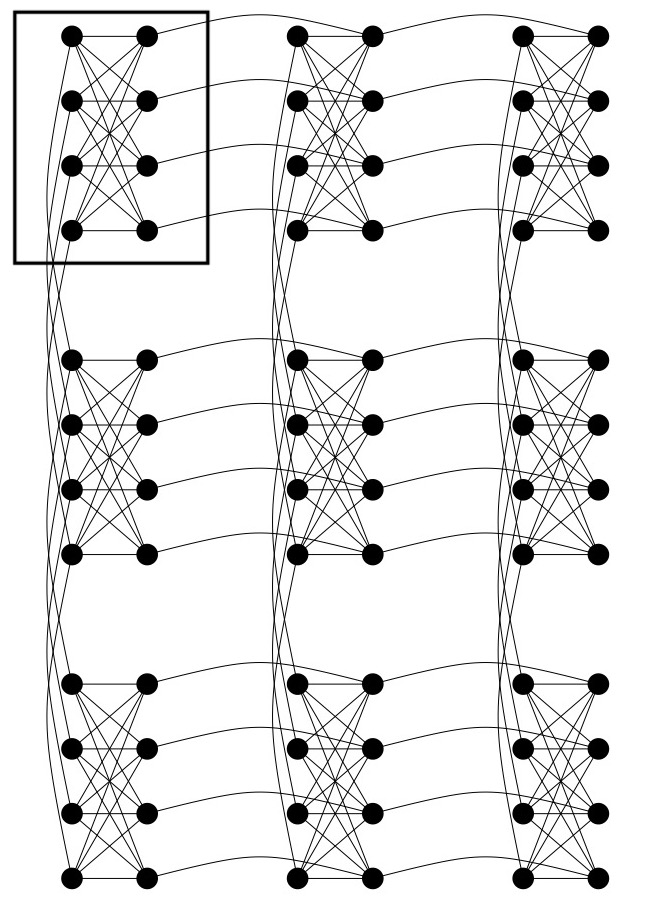} 
\caption{A unit cell of the Chimera graph and a subset of the D-Wave Chimera architecture showing how the unit cells are connected to each other. Each unit cell is a network of at most eight qubits in which each qubit is connected to at most six other qubits. All problems must be embedded onto this network, possibly using chaining for graph structures which do not fit naturally.}
\label{Fig:ChimeraGraph}
\end{figure}

A further technical issue is embedding. Though the model has so far been presented as a theoretical graphical network of qubits, this theoretical graph must be embedded onto the physical hardware of the D-Wave. The graph configuration chosen for the D-Wave hardware is called a Chimera structure, in which qubits are arranged in sparsely connected groups of at most six other qubits, called unit cells. A sample unit cell and the connected Chimera structure is shown in Figure  \ref{Fig:ChimeraGraph}.

\begin{figure} \centering
\includegraphics[height = 100mm]{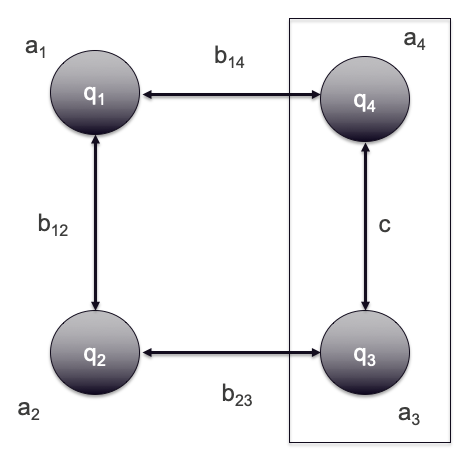}
\caption{The three-qubit system of Figure \ref{Fig:ThreeQubitSystem} mapped onto a Chimera graph using four qubits and a chain strength of $c$. The qubits $q_3$ and $q_4$ are treated as a single qubits and the value of $c$ should be chosen such that all low-energy solutions return $q_3 = q_4$. Using chains, large and complex graph structures can be embedded onto the Chimera structure used by the D-Wave hardware.}
\label{Fig:ChainDemonstration}
\end{figure}

The Chimera structure precludes the simple three-qubit system of Figure \ref{Fig:ThreeQubitSystem} from being solved directly on the D-Wave, as there is no set of three qubits all connected to each other on the Chimera structure. The graph must be embedded onto the Chimera structure to find the minimum energy of this system. To do this, multiple qubits must be chained together and the chain treated as if it were a single qubit. This is called a chain, and an embedding of the three-qubit system using four qubits and one chain is shown in Figure \ref{Fig:ChainDemonstration}. To chain qubits together, the problem must be first converted to Ising form if it is not already. In Ising form, $q_i q_j = 1$ when $q_i = q_j$ and $q_i q_j = -1$ when $q_i \ne q_j$. Then in place of an interaction term $b_{ij}$, a single chain strength $c$ is chosen for all connections between chained qubits such that, ideally, all low-energy solutions have qubits in a chain return the same value. It is possible to use multiple long chains to map complex graphical structures onto the Chimera structure. On the D-Wave machine, all chains must use the same chain strength $c$. D-Wave offers automated tools to find and perform an embedding and unembed the results automatically, currently available in C, Python, and Matlab.

The value of $c$ should be negative so that the total energy of the system is made smaller by having unbroken chains with $q_i = q_j$ for all $i, j$ in a chain. Within this constraint, it is clear that the magnitude of the chain strength $c$ must be chosen carefully. If $c$ is chosen too low in magnitude, then chains will be returned broken, with $q_i \ne q_j$. If $c$ is chosen too high in magnitude, then unnecessary rescaling will cause  the effects of noise introduced into the coefficients of the graphical network to degrade performance. Research has been performed on selecting an optimal choice for $c$, as in \cite{Rieffel2015}, but at the moment there is no solid criterion for choosing a value. It may be necessary to use the quantum annealer with multiple values of the chain strength $c$ in order to determine which is optimal for a given problem.

Embedding reduces the size of problems available to solve on the D-Wave, as a problem which takes tens of qubits in the theoretical graphical model may require hundreds or thousands of qubits when embedded on the Chimera structure. The necessity of embedding also magnifies the influence of the errors on the coefficients given in Equation (\ref{Eq:Noise}), adding hundreds or thousands of errors to the calculation. In general, reducing the size of the embedding will improve performance.

\subsection{Running a Program on the D-Wave}

Running a program on the D-Wave follows these steps:

\begin{enumerate}
\item Formulate the problem as a QUBO or Ising model, as in Equation (\ref{Eq:Energy}). 
\item Embed the graphical structure of the problem onto the Chimera graph in Figure \ref{Fig:ChimeraGraph}. If the original problem is in QUBO form, this will necessarily involve a conversion to the Ising model. This can be done either by hand or using the automated tools provided by D-Wave.
\item Specify any parameters of the anneal, such as the number of reads, anneal time, or other more advanced processing options.
\item Perform the quantum anneal and obtain the results.
\item Unembed the results from the Chimera structure to obtain a solution and energy for each read of the anneal. Analyzing them may involve simply choosing the solution returned with the lowest energy or performing some analysis over all returned solutions.
\end{enumerate}

Keeping all $a_i$ and $b_{ij}$ coefficients within hardware bounds and minimizing embedding generally improves performance of the quantum anneal. This paper will not focus on doing these things, instead using the machine as it is likely to be used by non-specialists: by formulating a problem into QUBO or Ising form, letting the automated tools provided by D-Wave perform the embedding and unembedding of the problem, and analyzing results. The goal is to use the hardware to solve practical optimization problems, many of which exist in statistics. To be clear, quantum computing should not be expected to outperform classical computing for years to come. The purpose is to develop algorithms and strategies for the use of quantum computers so that, when hardware developments do allow for quantum supremacy, statisticians are prepared to make full use of the technology.

\section{Maximum Likelihood Estimation}
\label{Sec:MaximumLikelihood}

\subsection{General Methodology}

One of the most common optimization problems in statistics is maximum likelihood estimation. Though the D-Wave has been used to solve, for example, linear least squares problems in \cite{Borle2019} and systems of polynomial equations in \cite{Chang2019}, most likelihood functions are non-polynomial. This section provides an algorithm which is, as far as the authors are aware, the first general purpose quantum annealing algorithm which can be applied to optimization of continuous functions.

Suppose that independent and identically distributed data $x_d$ for $d = 1, ..., n$ is available from some two-parameter distribution $f(x_d | \theta, \phi)$.  The maximum likelihood estimates are then 

\begin{equation}
(\hat{\theta}, \hat{\phi}) = \argmax_{(\theta, \phi)} \sum_d \ell(\theta, \phi | x_d)
\label{Eq:LogLikelihood}
\end{equation}

\noindent where $\sum_d \ell(\theta, \phi | x_d)$ is the standard log-likelihood.

In order to use the D-Wave, the log-likelihood function must be put in the form of Equation (\ref{Eq:Energy}).  Begin by writing the parameters $\theta$ and $\phi$ in binary. Suppose that a total of $n_q$ qubits in QUBO form are available for use, indexed by $q_i$ for $i = 1, 2, \dots, n_q$. Let the subscript $\theta$ on the index indicate that the qubit belongs to set of qubits which represent binary digits of $\theta$, and likewise the subscript $\phi$ on the index as the set of qubits which represent powers of $\phi$. The conversion back into decimal form is then

\begin{equation}
\hfill \theta = \sum_i 2^{p_{i_{\theta}}} q_{i_{\theta}} \hspace{1in} \phi = \sum_i 2^{p_{i_{\phi}}} q_{i_{\phi}} \hfill
\label{Eq:QubitPowers}
\end{equation}

\noindent where $p_{i_\theta}$ and $p_{i_\phi}$ are the powers of 2 to be used for the calculation of each parameter. For example, using five qubits for the calculation of $\theta$ with powers $p_{i_ \theta} = \{1,0,-1,-2,-3\}$ allows for $\theta$ to be any value between 0 and 3.875 with a numerical resolution of 0.125. This binary representation of numbers using qubits has previously been introduced in \cite{OMalley2016}.

Note that Equation (\ref{Eq:QubitPowers}) forces $\theta > 0$ and $\phi > 0$. It is easy to extend this to negative numbers by introducing a sign qubit as in  \cite{Borle2019}; however, this reduces the numerical resolution of the calculations. To maximize numerical resolution, only positive solutions were considered. In general, there may be information about the problem or the solution that can allow the range of powers of two in Equation (\ref{Eq:QubitPowers}) to be reduced, increasing the accuracy of the calculations.

Given the single datum $x_d$, expand the log-likelihood $\ell(\theta, \phi | x_d)$ around points $\theta_0$ and $\phi_0$ using a two-term Taylor series expansion.

\begin{multline}
\ell(\theta, \phi | x_d) \approx \ell(\theta_0, \phi _0 | x_d) + \ell_{\theta}(\theta_0, \phi _0 | x_d) (\theta - \theta_0) + \ell_{\phi}(\theta_0, \phi _0 | x_d) (\phi - \phi_0)\\ + \frac12 [\ell_{\theta \theta}(\theta_0, \phi _0 | x_d) (\theta - \theta_0)^2 + 2 \ell_{\theta \phi}(\theta_0, \phi _0 | x_d) (\theta - \theta_0) (\phi - \phi_0)   + \\ \ell_{\phi \phi}(\theta_0, \phi _0 | x_d) (\phi - \phi_0)^2 ]
\label{Eq:TaylorExpansion}
\end{multline}

\noindent where subscripts on the log-likelihood $\ell$ represent partial derivatives with respect to the notated parameter or parameters in the order given. A maximum power of two is chosen for the Taylor expansion because higher powers induce three-qubit interactions $c_{ijk} q_i q_j q_k$ or higher, which are not supported by the D-Wave hardware. 

Then plugging the representations in Equation ($\ref{Eq:QubitPowers}$) into Equation (\ref{Eq:TaylorExpansion}) wherever possible, expanding the polynomials, and collecting terms (with $q_i^2 = q_i$ since $q_i \in \{0, 1\}$), the following Equations are obtained:

\begin{equation*}
\ell(\theta, \phi | x_d) \approx \sum_i a_i q_i + \sum_{j > i} b_{ij} q_i q_j
\end{equation*}

\noindent where

\begin{equation}
a_i = 
\begin{dcases}
  -\sum_{d = 1}^n [  2^{p_{i_{\theta}}} \ell_{\theta}(\theta_0, \phi_0 | x_d) +  2^{2 p_{i_{\theta}} - 1} \ell_{\theta \theta}(\theta_0, \phi_0 | x_d) \\  \textrm{\hspace{0.5in}}  - 2^{p_{i_{\theta}}} \theta_0 \ell_{\theta \theta}(\theta_0, \phi_0 | x_d)- 2^{p_{i_{\theta}}} \phi_0 \ell_{\theta \phi}(\theta_0, \phi_0) ]& \textrm{$i$ is a qubit of $\theta$}\\
-\sum_{d = 1}^n [  2^{p_{i_{\phi}}} \ell_{\phi}(\theta_0, \phi_0 | x_d) +  2^{2p_{i_{\phi}} - 1} \ell_{\phi \phi}(\theta_0, \phi_0 | x_d) \\ \textrm{\hspace{0.5in}} - 2^{p_{i_{\phi}}} \phi_0 \ell_{\phi \phi}(\theta_0, \phi_0 | x_d)- 2^{p_{i_{\phi}}} \theta_0 \ell_{\theta \phi}(\theta_0, \phi_0) ] & \textrm{$i$ is a qubit of $\phi$}\\
\end{dcases}
\label{Eq:MLEa}
\end{equation}

\begin{equation}
b_{ij} = 
\begin{dcases}
-\sum_{d = 1}^n \left[2^{p_{i_{\theta}}+p_{j_{\theta}}} \ell_{\theta \theta}(\theta_0,\phi_0 | x_d)\right] & \textrm{$i,j$ are both qubits of $\theta$}\\
-\sum_{d = 1}^n \left[2^{p_{i_{\phi}}+p_{j_{\phi}}} \ell_{\phi \phi}(\theta_0,\phi_0 | x_d)\right] & \textrm{$i,j$ are both qubits of $\phi$}\\
-\sum_{d = 1}^n \left[2^{p_{i_{\theta}}+p_{j_{\phi}}} \ell_{\theta \phi}(\theta_0,\phi_0 | x_d)\right] & \textrm{$i,j$ are qubits of different parameters} \\
\end{dcases}
\label{Eq:MLEb}
\end{equation}

The set of qubits for the lowest-energy solution returned by the quantum annealer can be converted into numerical estimates $\hat{\theta}$ and $\hat{\phi}$ using Equation (\ref{Eq:QubitPowers}). Unfortunately, these $\hat{\theta}$ and $\hat{\phi}$ maximize only the approximated likelihood constructed using a two-term Taylor series expansion in Equation (\ref{Eq:TaylorExpansion}), not the full likelihood function in Equation (\ref{Eq:LogLikelihood}).  Because of this, the procedure is iterated by taking the $\hat{\theta}$ and $\hat{\phi}$ returned by the  quantum anneal, expanding the Taylor series around those values, and finding a new $\hat{\theta}$ and $\hat{\phi}$ values which maximize this function using a new quantum anneal. For most likelihood functions, which are well behaved, the values $\hat{\theta}$ and $\hat{\phi}$ will converge to the maximum likelihood estimates in Equation (\ref{Eq:LogLikelihood}). At the maximum likelihood estimates, expanding a Taylor series around $\hat{\theta}$ and $\hat{\phi}$ and then maximizing it will return those same $\hat{\theta}$ and $\hat{\phi}$ values. The ideas behind this iterative process are not new and not unique to quantum annealing, using the same principles as Newton and Raphson's original optimization method. A summary of the iterative process is given below.

\begin{enumerate}
\item Choose initial values $(\theta_0, \phi_0)$ such that $\ell(\theta_0, \phi_0 | x_d)$ and the first two derivatives at each point exist and are finite.
\item Expand the two-term Taylor series for $\ell(\theta_0, \phi_0 | x_d)$ around $(\theta_0, \phi_0)$, as in Equation (\ref{Eq:TaylorExpansion}).
\item Find lowest-energy values $(\hat{\theta}, \hat{\phi})$ using the quantum annealer with Equations (\ref{Eq:MLEa}) and (\ref{Eq:MLEb}) and binary representations in Equation (\ref{Eq:QubitPowers}). 
\item Take $(\hat{\theta}, \hat{\phi})$ as new expansion points $(\theta_0, \phi_0)$.
\item Repeat steps 2-4 until some stopping criterion is met.
\end{enumerate}

Because of the noise inherent in the system and the approximate nature of quantum annealing, a stopping criterion based on a tolerance level will  fail to converge unless set large. From Equations (\ref{Eq:MLEa}) and (\ref{Eq:MLEb}), the values $a_i$ and $b_{ij}$ values which are smallest in magnitude are those of the least significant qubits due to the influence of the $2^{p_{i}}$ terms. These are precisely the ones which contribute the least to the total energy of the system and that will be most affected by the noise in the model inputs, as described in Equation (\ref{Eq:Noise}). It is unlikely that the method will fix upon the exact values which minimize the energy of the system, but will vary randomly in the least significant qubits.

\subsection{Example}

A simple test case for this algorithm is to estimate the parameters of a $N(\theta, \phi^2)$ distribution from a random sample of data. Using parameters $\theta = 0.5$ and $\phi = 1$, a random sample  $x_d = [ -2.296, -0.216, -0.082,  0.231, 1.127,  1.164, 1.189, 1.236,  1.272,\\  1.373]$ was generated. The maximum likelihood estimates for this data set are $\hat{\theta} = \bar{x} = 0.4998$ and $\hat{\phi} \approx 1.093$.

\begin{figure} \centering
\includegraphics[height = 60mm]{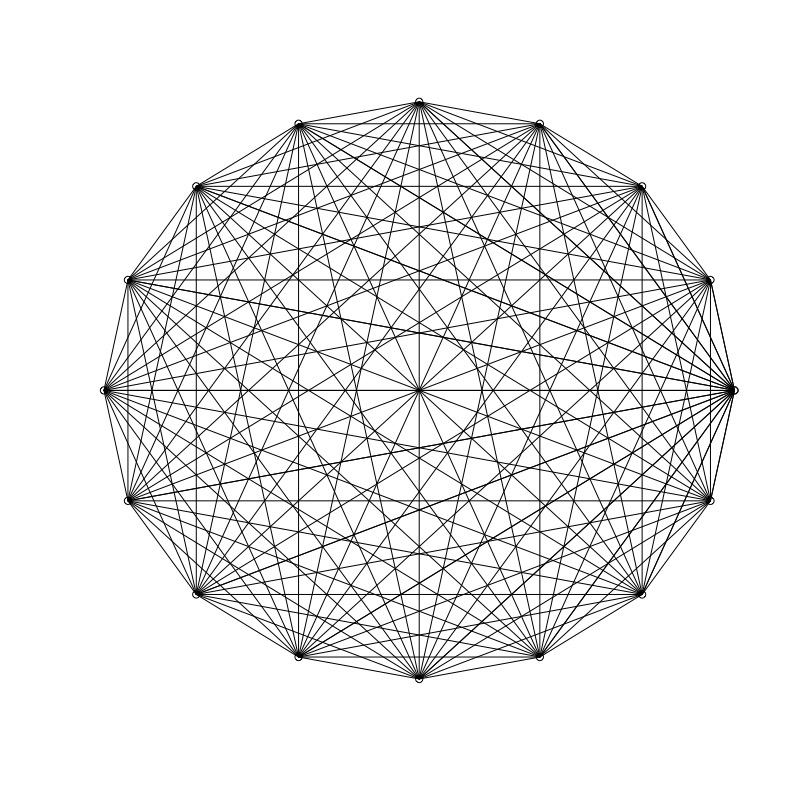}
\includegraphics[height =60mm]{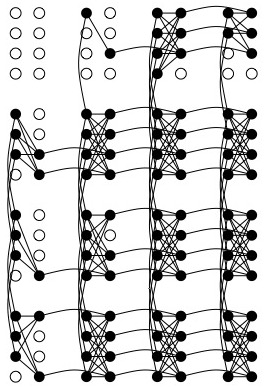}
\caption{The graph of all qubits required for the maximum likelihood annealing step is a complete $K_{n_q}$ graph. On the left, a complete $K_{16}$ graph representing all the connections between the qubits of $\theta$ and $\phi$ in the example. On the right, the embedding of the $K_{16}$ graph onto the Chimera hardware of the D-Wave. The Chimera graph is well-suited for this complete graphical network.}
\label{Fig:MLQubitGraph}
\end{figure}

A total of 10 iterations of the algorithm ran on the LANL D-Wave 2X computer, using D-Wave software embedding tools with chain strength $c = -5.0$. Eight qubits were used for each variable, with powers $p_{i_{\theta}} = p_{i_{\phi}} = \{1,0,-1,-2,-3, -4,\\ -5, -6, -7\}$, for sixteen qubits total. Following Equations (\ref{Eq:MLEa}) and (\ref{Eq:MLEb}), each qubit interacts with all other qubits, forming a complete $K_{16}$ graph for the problem. This graph, and its resulting embedding onto the Chimera structure, are shown in Figure \ref{Fig:MLQubitGraph}. The Chimera network is efficient at embedding this complete graph, requiring approximately only 90 qubits total.

Starting values were $\theta_0 = 0$ and $\phi_0 = 1$.  The minima and energy at each iteration are given in Table \ref{Tab:QNormal}. The chosen powers of 2 give a maximum numerical resolution of $2^{-7} =  0.0078125$, and this is apparent in the results. After an initial period of convergence, the method samples randomly around the correct maximum likelihood estimates as closely as the numerical resolution allows. In fact, by the fourth iteration the algorithm converged as closely as possible given the numerical resolution. The results of Table \ref{Tab:QNormal} show that the method is performing well on this simulated data set.

\begin{table}[h]
    \centering
    \begin{tabular}{|c|c|c|c| c |} \hline
        Iteration & $\hat{\theta}$ & $\hat{\phi}$ & Energy & $\ell(\theta, \phi | x_d)$ \\ \hline
	1 & 0.5078125 & 0.9765625 & -423.439 & -15.218\\
	2 &  0.5 & 1.0625 & -422.93 & -15.089\\
	3 & 0.515625 & 1.0859375 & -420.865 & -15.082  \\
	4 &  0.5 & 1.09375 & -420.434 & -15.080\\
	5 &  0.5 & 1.0859375 & -420.283 & -15.081\\
	6 &  0.4765625 & 1.09375 & -420.42 & -15.082\\
	7 &  0.53125 & 1.09375 & -420.263 & -15.084\\
	8 & 0.484375 & 1.09375 & -420.308 & -15.081\\
	9 & 0.5 & 1.09375 & -420.272 & -15.080\\
	10 &  0.5 & 1.0859375 & -420.283 & -15.081 \\        \hline
    \end{tabular}
    \caption{Ten iterations of the D-Wave quantum annealer, with starting values $\theta_0 = 0$ and $\phi_0 = 1$. Eight qubits were used for each parameter, sixteen total, with a maximum numerical resolution of $2^{-7} = 0.0078125$. At each iteration, the two-term Taylor series for the log-likelihood expanded around $\hat{\theta}$ and $\hat{\phi}$ from the previous iteration was maximized using the quantum annealer. The results show that the quantum annealing algorithm is performing well on this data set. The energy is not simply a function of the log-likelihood $\ell(\theta, \phi | x_d)$ due to both the the necessity of using a two-term Taylor series approximation of the log-likelihood and embedding introducing many more qubits into the annealing procedure, each of which contributes to the energy total.} 
    \label{Tab:QNormal}
\end{table}

\subsection{Method Criticisms}

There is much work to do in the general use of quantum annealing for maximum likelihood optimization, and this method is only a step. Constructing and maximizing a quadratic surface at each iteration leads only to the nearest local minima rather than the global minima, in many ways defeating the purpose of quantum annealing. One possible solution to this issue is including more terms in the Taylor expansion of Equation (\ref{Eq:TaylorExpansion}). This will create interactions of three or more qubits, but these can be handled by two-qubit interactions combined with penalty functions. In fact, through the use of penalties to represent higher-order qubit interactions, any function which has a Taylor series representation has a theoretical corresponding QUBO representation, though such representation will necessarily be truncated in practice. A potential approach is to attempt to estimate said QUBO representation using the function itself as an objective, as in machine learning.  Lastly, the method here is for two parameters, but it may be expanded to any number of parameters given the appropriate dimension Taylor series expansion and algebraic manipulations. All of these improvements require more qubits or the ability to form larger networks between them. Though not possible on current hardware, they may be of use on quantum annealers of the future. 


\section{Experimental Design}
\label{Sec:NQueens}


A more realistic use of quantum annealing is for problems which currently rely on techniques such as simulated annealing or genetic algorithms, such as the case of experimental design. Finding an optimal design generally scales exponentially with the size or dimensionality of the problem, and quantum annealing may provide a polynomial increase in computational speed for high dimensional problems.

Suppose an $N \times N$ Latin hypercube is desired. Current methods for finding a space-filling Latin hypercube use simulated annealing to find the design which is maximin, as in \cite{Morris1995}.  

\subsection{General Methodology}

\begin{figure} \centering
\includegraphics[height = 60mm]{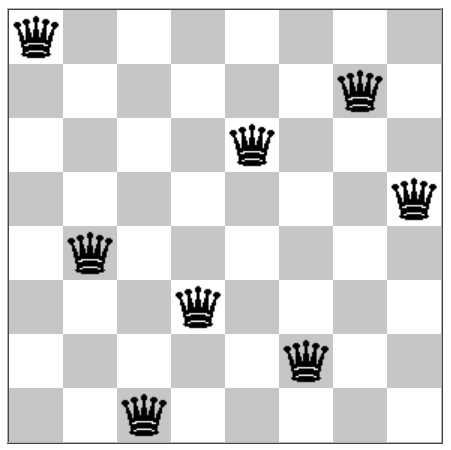}
\includegraphics[height =60mm]{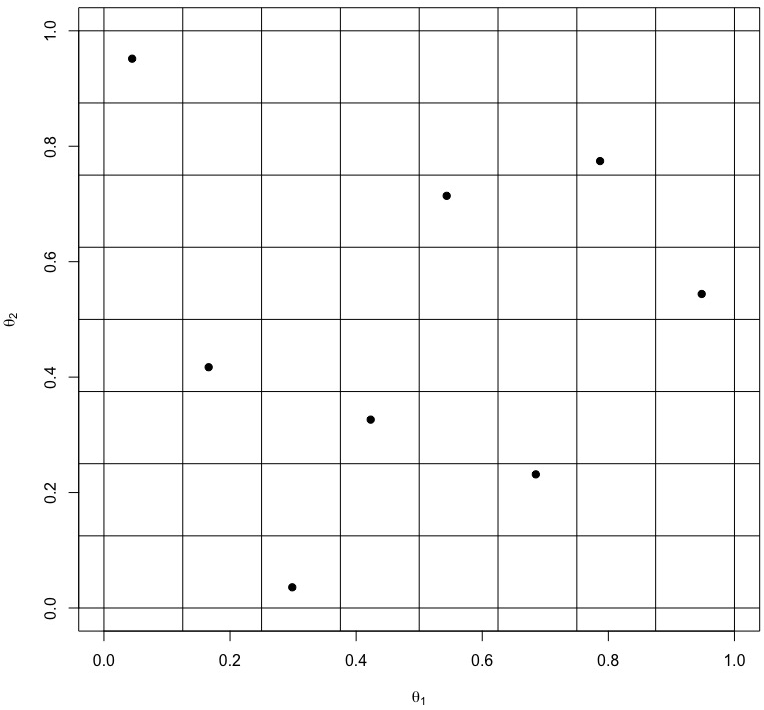}
\caption{One solution to the 8-Queens problem and a corresponding Latin hypercube generated from the design.}
\label{Fig:8Queens}
\end{figure}

Suppose that instead of a maximin or some other distance-based criterion, a design is desired for which there is one observation within each row, column, and diagonal of the design matrix. The row and column requirements force a Latin hypercube while the diagonal requirements force some degree of space-filling, or may be desired for its own properties depending on the particular problem at hand. This is the classic $N$-queens problem of placing $N$ queens on an $N \times N$ chessboard such that no queen is directly attacking another queen.  Though found in many references, \cite{Mandziuk1992} gives a more thorough description of the problem including an equation for the energy in the form of Equation (\ref{Eq:Energy}).  An example solution to the 8-queens problem and corresponding Latin hypercube is shown in Figure \ref{Fig:8Queens}.

Each square on the grid can be represented in QUBO form by a single qubit, $q_i, i = 1, 2, \dots, N^2$, where $q_i = 0$ represents no design point in the square and $q_i = 1$ represents a design point in the square. The particular labeling is not important, so long as qubits can consistently be identified as belonging to the same row, column, and diagonal. An example layout of a $4 \times 4$ design is shown in Figure \ref{Fig:4by4} which labels the squares sequentially in order from left to right and top to bottom.

\begin{figure} \centering
\begin{tikzpicture} \large
\draw[step=0.5cm,color=gray, scale = 2] (-1,-1) grid (1,1);
\node at (-1.5,+1.5) {$q_1$};
\node at (-0.5,+1.5) {$q_2$};
\node at (+0.5,+1.5) {$q_3$};
\node at (+1.5,+1.5) {$q_4$};
\node at (-1.5,+0.5) {$q_5$};
\node at (-0.5,+0.5) {$q_6$};
\node at (+0.5,+0.5) {$q_7$};
\node at (+1.5,+0.5) {$q_8$};
\node at (-1.5,-0.5) {$q_9$};
\node at (-0.5,-0.5) {$q_{10}$};
\node at (+0.5,-0.5) {$q_{11}$};
\node at (+1.5,-0.5) {$q_{12}$};
\node at (-1.5,-1.5) {$q_{13}$};
\node at (-0.5,-1.5) {$q_{14}$};
\node at (+0.5,-1.5) {$q_{15}$};
\node at (+1.5,-1.5) {$q_{16}$};
\end{tikzpicture}
\caption{Layout of qubits used for generation of a Latin hypercube on a regular 4 x 4 grid. The particular layout of qubits is not important, so long as it is possible to identify which qubits belong to the same row, column, and diagonal.}
\label{Fig:4by4}
\end{figure}

Letting $q_i$ and $q_j$ represent two qubits such that $i \ne j$, the energy of the system in the form of Equation (\ref{Eq:Energy}) is

\begin{equation}
a_i = -2 \textrm{ \hspace{0.5in} } b_{ij} = 
\begin{dcases}
2 & \textrm{ for $i, j$ in the same row}\\
2 & \textrm{ for $i, j$ in the same column}\\
1 & \textrm{ for $i, j$ in the same forward diagonal}\\
1 & \textrm{ for $i, j$ in the same backward diagonal}\\
0 & \textrm{ otherwise }\\
\end{dcases}
\label{Eq:NQueensEnergy}
\end{equation}

\noindent The minimum possible energy for this system is $Energy = -4N$, reached at the solution where no two qubits in the same row, column, or diagonal are both set to $q_i = 1$, as in Figure \ref{Fig:8Queens}. Note that this system is not unique in defining the problem, as different choices of $a_i$ and $b_{ij}$ in Equation (\ref{Eq:NQueensEnergy}) will still yield the N-queens solution as the lowest-energy design but with differing energies for other designs. The particlar values in Equation (\ref{Eq:NQueensEnergy}) penalize violation of the Latin hypercube criterion moreso than violation of the diagonal criterion and were chosen in an attempt to mitigate the risk of failing to obtain the lowest-energy design. Manipulating the distribution of energies in order to emphasize certain classes of designs while still maintaining the preferred design as lowest-energy is perhaps an interesting area for further research.

\subsection{Example}

The D-Wave 2000Q minimized the energy function in Equation (\ref{Eq:NQueensEnergy}) for multiple sized Latin hypercubes. Unfortunately, this set of connections does not readily lend itself to embedding on the Chimera graph as strongly as the complete $K_{16}$ graph of Figure \ref{Fig:MLQubitGraph}. Using the embedding software provided by D-Wave and the D-Wave 2000Q computer, embedding the 36 qubits of a 6 x 6 hypercube uses approximately 400 total qubits, while embedding the 64 qubits of an $ 8 \times 8$ hypercube requires roughly 1400 qubits. In fact, the 8 x 8 hypercube is the largest which, with embedding, fits on the D-Wave 2000Q.

\begin{figure} \centering
\includegraphics[height = 60mm]{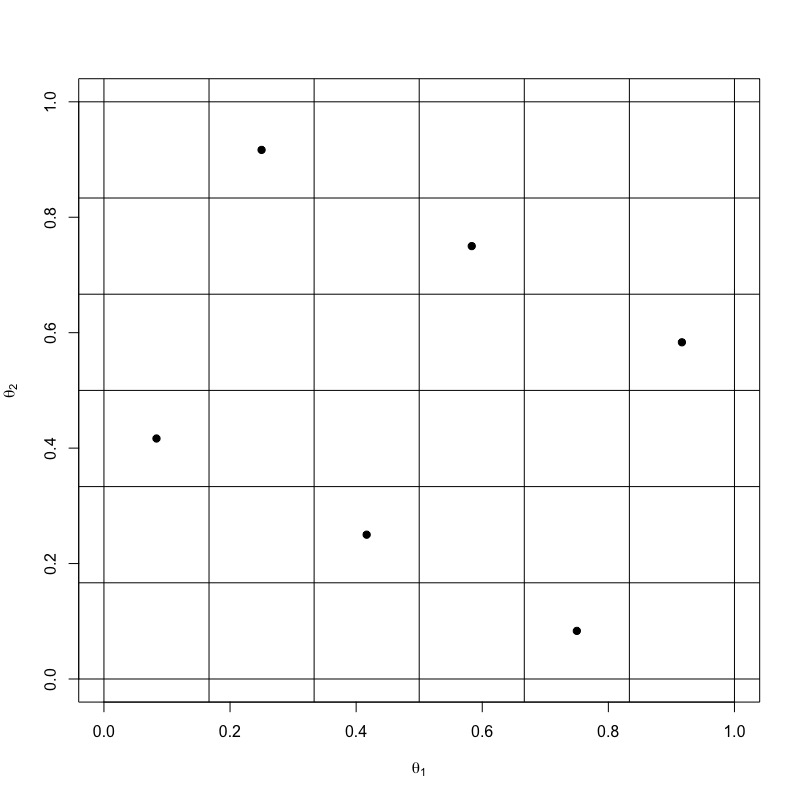}
\includegraphics[height =60mm]{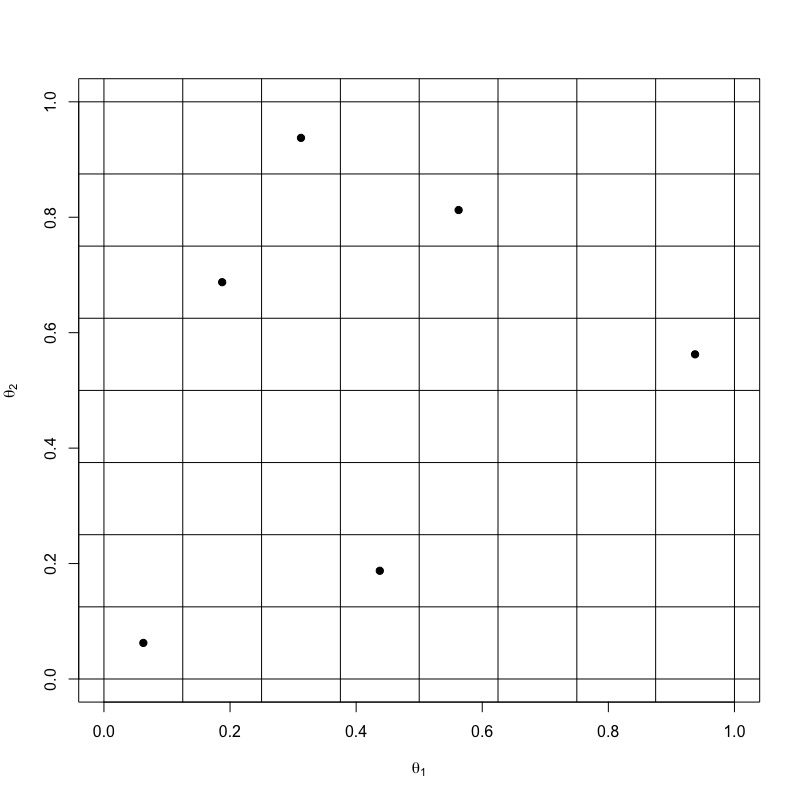}
\caption{A 6 x 6  and an 8 x 8 Latin hypercube generated using quantum annealing, following the N-queens energy function. The 6 x 6 is the largest Latin hypercube for which the lowest-energy solution could be found, while the 8 x 8 is the largest hypercube that would fit onto the D-Wave 2000Q machine. The performance degrades as the problem size increases. Noticeably, the 8 x 8 hypercube does not even feature eight design points.}
\label{Fig:LatinHypercubes}
\end{figure}

For each hypercube, a chain strength of $c = -5.0$ was used. The largest sized hypercube for which a minimum energy solution could be found was a 6 x 6, using a 99 $\mu s$ anneal and 10000 reads, while using the same parameters for an 8 x 8 hypercube did not yield a good solution - the lowest-energy solution returned failed to place eight total observations. Both are shown in Figure \ref{Fig:LatinHypercubes}. This analysis was repeated using a variety of chain strengths and annealing times, but there did not appear to be a significant improvement in performance.  

Solutions to the $N$-queens problem are not unique. There are multiple hypercubes which achieve the minimum energy of $-4N$, many of which are simple rotations or reflections of existing hypercubes. Due to noise in the hardware as in Equation (\ref{Eq:Noise}), however, the D-Wave will tend towards one particular solution above all others. This tended to be one particular solution within a set of anneals but different solutions between different sets anneals, possibly due to the effect of temporal correlation in the noise.

In theory, the weights given in Equation (\ref{Eq:NQueensEnergy}) can be rescaled to induce a large difference between the lowest and second-lowest energy solution, so that the system may more easily remain in the lowest-energy state throughout the entire anneal. In practice, the rescaling of coefficients performed by the D-Wave prevents this. Furthermore, the large number of qubits required increases the number of points at which errors introduced through thermal or chaining effects may affect the final result. Allowing for a wider range of input values, or allowing for additional connections between qubits, may help increase the potential problem size.

\section{Matrix Inversion}
\label{Sec:MatrixInversion}

General models similar to the form of Equation (\ref{Eq:Energy}) exist, and a number of problems have been already been placed into said formats in a manner that is easy to convert to the QUBO or Ising model of the D-Wave. For example,  \cite{Jang1988} presents matrix optimization using a Hopfield network, a form of recurrent artificial neural network strikingly similar to that of the Ising-spin model given in Equation (\ref{Eq:Energy}), and which is easily adapted for use in quantum annealing. Matrix inversion is of particular important in statistics, as it often serves as the bottleneck in fitting statistical models such as Gaussian process regressions. Any method that would decrease computational time at this bottleneck would greatly increase the applicability of such techniques.

It should be noted that this is not the only approach to using the D-Wave machine to perform matrix inversion. \cite{Rogers2019} presents a similar method for inverting matrices using the D-Wave hardware, though approaching the problem as solving a set of linear Equations rather than the Hopfield network of \cite{Jang1988}. Other matrix include nonnegative binary matrix factorization in \cite{OMalley2018}.

\subsection{General Methodology}
Define $A$ as an $n \times n$ matrix and $V$ as its corresponding inverse, respectively. Then from \cite{Jang1988}, column $k$ of $V$ may be found individually by minimizing the energy Equation

\begin{equation}
E_k = \left( \sum_{\ell = 1}^n A_{1\ell}V_{\ell k} \right)^2 + \ldots +  \left( \sum_{\ell = 1}^n A_{k \ell }V_{\ell k} - 1\right)^2 +  \left( \sum_{\ell = 1}^n A_{n \ell}V_{\ell k} \right)^2 
\label{Eq:ColumnEnergy}
\end{equation}

Writing the elements $V_{rk}$ as sums of powers of two times qubits in QUBO form, similarly to Equation (\ref{Eq:QubitPowers}), gives

\begin{equation}
V_{rk} = \sum_{\ell = 1}^{n_{rk}} 2^{p_{\ell_{rk}}} q_{p_{\ell_{rk}}} = 2^{p_{1_{rk}} }q_{1_{rk}} + \ldots + 2^{p_{n_{rk}}} q_{n_{rk}} 
\label{Eq:VCQubits}
\end{equation}

\noindent where $n_{rk}$ is the number of qubits used in the binary representation of $V_{rk}$ and once again, $\ell_{rk}$ indicates that the qubit represents a power of $V_{rk}$. Expanding and collecting terms, the energy Equation (\ref{Eq:ColumnEnergy}) for column $k$ of V can be rewritten in the form of Equation (\ref{Eq:Energy}) as\\


\begin{equation*}
a_i =2^{2 p_{i_{rk}}} \alpha_{r} -  2^{p_{i_{rk}}+1} A_{kr}  \textrm{\hspace{0.1in}$i$ is a qubit of $V_{rk}$} \\
\end{equation*}

\begin{equation*}
b_{ij} = 
\begin{dcases}
2^{p_{i_{rk}} + p_{j_{rk}}+1} \alpha_{r}  & \textrm{$i, j$ are qubits of the same $V_{rk}$}\\
2^{p_{i_{rk}} + p_{j_{rk}}+1} \beta_{r_i, r_j}  & \textrm{$i, j$ are qubits of different $V_{rk}$}\\
\end{dcases}
\end{equation*}


\noindent where the quantities $\alpha_r$ and $\beta_{r_1, r_2}$ are given by\\

\begin{gather*}
\alpha_{r} = \sum_{\ell = 1}^n A_{\ell r}^2 \\
\beta_{r_1, r_2} = \sum_{\ell = 1}^n A_{\ell r_1} A_{\ell r_2} 
\end{gather*}

The quantities $\alpha_i$ and $\beta_{ij}$ only need to be calculated once, before the actual quantum anneal. In fact, nearly the entire matrix $Q$ (as in Equation (\ref{Eq:QUBOMatrix})) may be precomputed, as only the subtraction of the $A_{kr}$ elements in the calculation of the $a_i$ differs between minimizations for different columns $k$. Unfortunately, creating the entire set of $\beta_{ij}$ requires calculating $\frac12 n(n-1)$ objects each requiring a sum of $n$ terms, precluding this method from improving on the fastest classical matrix inversion algorithms. This time could be reduced by parallel computation, both in classical calculation of the $\alpha_i$ and $\beta_{ij}$ and, supposing one had access to multiple quantum annealers, minimization of the energy function in Equation (\ref{Eq:ColumnEnergy}).

\subsection{Example}

This method was tested on matrices of multiple sizes with known inverses consisting entirely of positive entries. An example $3 \times 3$ matrix $A$ and corresponding inverse $V$ are shown below.

\begin{center}

\[
\begin{bmatrix}
1.344 & 0.418 & -0.935\\
-1.018 & 1.095 & -0.250\\
0.277 & -0.384 & 0.755\\
\end{bmatrix}
\times 
\begin{bmatrix}
0.613 & 0.037 & 0.772\\
0.586 & 1.068 & 1.080\\
0.074 & 0.531 & 1.592\\
\end{bmatrix}
=
\begin{bmatrix}
1 & 0 & 0\\
0 & 1 & 0\\
0 & 0 & 1\\
\end{bmatrix}
\] 
\hspace{0.65in} $A$ \hspace {1.7in} $V$ \hspace{1.0in} $I_{3 \times 3}$

\end{center}

Using six qubits per element of $V_{rk}$ - 18 qubits total for the column - the estimated inverse $V$ of the above matrix $A$ was found using the D-Wave 2000Q. With embedding, a total of 125 qubits were used. Using 2500 reads,  a default annealing time, and $c = -10$ chain strength, the resulting estimate is shown below.

\begin{center}
\[
\begin{bmatrix}
1.344 & 0.418 & -0.935\\
-1.018 & 1.095 & -0.250\\
0.277 & -0.384 & 0.755\\
\end{bmatrix}
\times 
\begin{bmatrix}
0.625 & 0.0 & 0.75\\
0.5 & 1.0625 & 1.1875\\
0.0 & 0.5625 & 1.6875\\
\end{bmatrix}
=
\begin{bmatrix}
1.049 & -0.082 & -0.073\\
-0.088 & 1 .023 & 0.116\\
-0.019 & 0.016 & 1.025\\
\end{bmatrix}
\]

\hspace{0.4in} $A$ \hspace {1.6in} $V$ \hspace{1.7in} $I_{3 \times 3}$
\end{center}


The quantum annealer was not able to find the inverse $V$ precisely, though this is in part due to the limited numerical resolution. The diagonal elements of the estimated inverse are reasonably close to the true values, with $V_{33}$ being the most off at 6\% relative error. The off-diagonal elements show larger relative errors.

In practice, $3 \times 3$ is the largest matrix size for which an estimated inverse can be consistently obtained that is recognizable as approximating the true inverse. The D-Wave 2000Q can handle up to an $8 \times 8$ matrix $A$, depending on the number of bits used for each column element $V_{rk}$; however, this does not yield a good estimate of the inverse matrix. Note that the discretization used in Equation (\ref{Eq:VCQubits}) can have a significant effect upon the final result. The energy measures the distance of the product of $A$ and the estimate of $V$ from the identity matrix, not the distance of the estimate of $V$ from the true $V$. Because of this, moving an individual element of the estimate $V_{rk}$ closer to its true value while changing nothing else can produce a larger energy in terms of the energy function in Equation (\ref{Eq:ColumnEnergy}). Multiple qubits must be adjusted simultaneously, providing a deceptively good test case for the D-Wave hardware. Multi-qubit tunneling provides another instance of a bottleneck which may have prevented the anneal from reaching the lowest-energy state, in addition to the aforementioned errors introduced by chaining and thermal effects. Once again, alternate chain strengths and annealing times did not significantly affect the results. 

\section{Conclusion and Future Research Directions}

Though quantum computing has arrived in at least one form, it is still not accurate enough to be incorporated into regular use in statistics. Qubit coherence remains an issue for quantum computers of all types, and theoretical research is ongoing into which types of problems may provide a speedup over classical computation using quantum annealing by analyzing how the gap between lowest and second-lowest energy states scales with problem size. The D-Wave, though more user-friendly than universal quantum computers, forces all problems to be formatted in the Ising or QUBO model of Equation (\ref{Eq:Energy}) - not a trivial task. The machine itself suffers from hardware noise which may mask the true solution and the process of embedding reduces the scale of problems available. The combined effects of these issues were shown in Sections \ref{Sec:NQueens} and \ref{Sec:MatrixInversion}, where larger problems could be embedded onto the machine, but suitable solutions could not be found. The effects of noise and difficulty of annealing with hundreds or thousands of qubits leads to solutions which are easily identifiable as incorrect but for which the correct solution can not be enforced, as in the solution to the N-queens problem with too few queens.  Furthermore, while the machine has been proven to utilize quantum tunneling effects not available to classical solver, it is still unclear whether these effects are even useful in solving the problem or scalable as the problem size grows. 

Those qualifications aside, the D-Wave is still potentially relevant for problems of the form in Equation (\ref{Eq:Energy}). This paper has shown maximum likelihood estimation, experimental design generation, and matrix inversion on the D-Wave. Of these, experimental design generation arguably shows the most promise for future research using quantum computing, as it relies on simulated annealing techniques for finding designs which satisfy an optima criterion - a problem for which the quantum annealing is well suited.


This paper will hopefully serve as inspiration for further research in quantum computing. Though noisy, hardware is available for current research into quantum computing. Proof of concept is here, and what remains  is for quantum annealers to find ways to make these current drawbacks in implementing the computational model in hardware more faithful so that the solutions are reliable.  It will be exciting to observe how this new technology is used to broaden computational possibilities in the field of statistics.

\vfill



\cleardoublepage


\bibliographystyle{jasa}

\bibliography{bibliography}

\end{document}